\documentclass[twocolumn,showpacs,preprintnumbers,amsmath,amssymb]{revtex4}

\usepackage{graphicx}
\usepackage{dcolumn}
\usepackage{bm}

\begin{document}

\preprint{APS/123-QED}

\title{Single-quasiparticle stability and \\
quasiparticle-pair decay in YBa$_2$Cu$_3$O$_{6.5}$ (Ortho II)}

\author{N. Gedik}
\email{gedik@socrates.berkeley.edu}
\author{P. Blake}
\author{R.C. Spitzer}
\author{J. Orenstein}
\affiliation{Physics Department, University of California,
Berkeley and \\ Materials Science
Division, Lawrence Berkeley National Laboratory, Berkeley, CA 94720}%

\author{Ruixing Liang}
\author{D.A. Bonn}
\author{W.N. Hardy}

\affiliation{Department of Physics and Astronomy, University of British Columbia, Vancouver, British Columbia, Canada V6T 1Z1 }%

\date{\today}

\begin{abstract}
We report results and analysis of time-resolved photoinduced
reflectivity experiments on the cuprate superconductor
YBa$_2$Cu$_3$O$_{6.5}$. The sample, which has $T_c$=45 K, was
characterized by a high degree of purity and Ortho II ordering.
The change in reflectivity $\Delta R$ was induced and probed using
pulses of 100 femtosecond duration and photon energy 1.55 eV from
a Ti:Sapphire laser. We provide a detailed picture of the decay
rate $\gamma$ of $\Delta R$ as a function of temperature $T$ and
pump intensity $I$. At low $T$, $\gamma$ decreases linearly with
decreasing $I$, extrapolating to nearly zero in the limit that $I$
tends to zero. At higher temperature $\gamma$ has the same linear
dependence, but with nonzero limit as $I\rightarrow0$. In the
interpretation of these results we assume that $\Delta R$ is
proportional to the nonequilibrium quasiparticle density created
by the laser. From an analysis of the $\gamma$ vs. $I$ we estimate
$\beta$, the coefficient of proportionality relating the
quasiparticle decay rate to the density. The intercept of $\gamma$
vs. $I$ yields the thermal equilibrium quasiparticle decay rate.
In a discussion section, we argue that the quasiparticles induced
by the laser occupy primarily states near the antinodal regions of
the Brillouin zone. We explain the divergence of the lifetime of
these particles as $T$ and $I$ both tend to zero as a consequence
of momentum and energy conservation in electron-electron
scattering. Next, we discuss the significance of the measured
value of $\beta$, which is $\approx$ 0.1 cm$^2$s$^{-1}$. We point
out that the natural unit for $\beta$ in a two-dimensional
superconductor is $\hbar/m^\ast$, and define a dimensionless
constant $C$ such that $\beta\equiv C\hbar/m^\ast$. If the decay
process is one in which quasiparticles return to the condensate
with emission of a phonon, then $C$ is a measure of the
electron-phonon interaction. Alternatively, expressing the
marginal Fermi liquid scattering in the normal state in terms of
an effective $\beta$ implies $C=1/\pi$, which is in excellent
agreement with the experimentally determined value in the
superconducting state.
\end{abstract}

\pacs{74.25.Gz, 78.47.+p}
\maketitle

\section{\label{sec:level1}Introduction}

A special property of the cuprate superconductors is that the
energy required for the creation of a quasiparticle depends on the
direction of its momentum \cite{shen}.  The creation energy is
zero for momenta in the 'nodal' direction, oriented at 45$^\circ$
relative to the Cu-O bond. The most energetically expensive
quasiparticles are the 'antinodal' ones, whose momenta are nearly
parallel to the bond.  Because they feel the pairing interaction
most strongly, their properties may hold the key to high-$T_c$
superconductivity.  Unfortunately, their tendency to form strong
pairs makes them difficult to study. In thermal equilibrium the
population of quasiparticles is overwhelmingly dominated by the
low energy nodal ones.  As a result, transport measurements
performed in equilibrium, such as microwave \cite{hosseini} and
thermal \cite{ong} conductivity, are insensitive to antinodal
quasiparticles.

A potentially powerful approach to studying the interactions of
antinodal quasiparticles at low temperature is to create a
nonthermal population by external excitation. By probing the
relaxation of the nonthermal population to the ground state, one
may hope to learn about interactions between quasiparticles that
would not normally be present at low temperature. Time-resolved
optical techniques are ideally suited to creating a nonequilibrium
density of quasiparticles and measuring their subsequent
relaxation. These techniques are based on mode-locked lasers that
produce pulses of 10-100 fs duration with a wide range of energy
per pulse, center wavelength, and repetition rate.  Such
measurements are performed in a 'pump-probe' mode in which a beam
of pulses is split in two. The pump pulse creates the
nonequilibrium state while the probe pulse senses the change in
the optical response of the medium due to the nonequilibrium
population.  The time delay between the two pulses is controlled
continuously and accurately by varying the optical path length
difference between the two beams. Measuring the transmission or
reflection of the probe as a function of time delay gives
information about the return to equilibrium after the pulsed
excitation.

There have been several time-resolved optical measurements
performed on the YBa$_2$Cu$_3$O$_{7-x}$ (YBCO) system of cuprate
superconductors. The earliest work \cite{han, eesley, chwalek,
brorson} reported the change in the reflectivity ($\Delta R$) of a
1.5 eV probe due to photoexcitation at the same energy. The
measurements showed that in the normal state $\Delta R$ is small
and decays very rapidly. Upon cooling below $T_c$, $\Delta R$
increases rapidly, and its decay rate decreases. These results
suggested that carrier thermalization and/or recombination proceed
rapidly in the normal state, but are strongly impeded by the
opening of the superconducting gap.

Subsequent measurements provided a more detailed picture of the
magnitude and decay rate of $\Delta R$ as a function of carrier
concentration \cite{demsar}.  In underdoped samples $\Delta R$ is
readily detectable in the normal state. The decay rate slows with
cooling, suggesting a correlation between the relaxation rate of
the nonequilibrium state and the appearance of the pseudogap. It
was reported that the relaxation time is insensitive to $T$ in the
superconducting state, with the exception of a peak near $T_c$.
The results were interpreted in terms of an excited state in which
quasiparticles and phonons rapidly reach quasiequilibrium.  The
decay rate was conjectured to reflect the thermalization of
nonequilibrium phonons.

Recently, measurements of $\Delta R$ at 1.5 eV were reported on an
untwinned single crystal of YBa$_2$Cu$_3$O$_{6.5}$ Ortho II
\cite{segre}. (The designation Ortho II refers to the macroscopic
ordering of the atomic layer containing the Cu-O chains.  In the
Ortho II phase the chains alternate between fully occupied, and
entirely unoccupied, by O atoms). In these measurements the range
of pump intensity was extended to nearly two orders of magnitude
smaller than used in earlier measurements. It was shown that while
the $T$ dependence of the decay rate is weak when the intensity,
$I$, is large, it becomes extremely strong as $I$ is reduced. In
fact, the decay rate appeared to vanish as $T$ and $I$ both tend
to zero.  It was suggested that the strong $I$ dependence of the
decay rate is not consistent with a picture in which the
quasiparticles and phonons reach quasiequilibrium and an alternate
mechanism involving the pairwise scattering of quasiparticles was
suggested.

In this paper, we present further measurements and analysis of
time-resolved photoinduced reflectivity in YBCO Ortho II. The
experimental apparatus and sample characterization are described
in Section II. In Section III we introduce the Rothwarf-Taylor
(RT) equations, which provide a phenomenological framework for
interpreting nonequilibrium dynamics in superconductors. Section
IV presents measurements of the decay of the photoinduced
reflectivity as a function of time, temperature, and pump beam
intensity.  The essential experimental finding, as in Ref.
\cite{segre}, is that the characteristic decay time diverges as
both $T$ and $I$ approach zero. However, we present several
results beyond those already reported.  We present and analyze the
time dependence of the transient reflectivity, showing that it is
described well by the pairwise scattering of quasiparticles.
Second, we report improved measurements of the asymptotic value of
the decay rate in the limit that $I$ goes to zero. According to
the RT equations, the low-intensity limit is a direct measure of
$\gamma_{th}$, the quasiparticle recombination rate in thermal
equilibrium. Through improvements in sensitivity we were able to
measure $\gamma_{th}$ with more than one order of magnitude
greater precision than previously. Finally, in Section V we
present analysis and interpretation of the experimental results.
We argue that the quasparticles that give rise to $\Delta R$ are
antinodal in character and interpret the decay rate as a measure
of the strength of antinodal quasiparticle interactions.

\section{\label{sec:level1}Experimental methods}

Pump and probe measurements were performed on a mechanically
detwinned single crystal of YBa$_2$Cu$_3$O$_{6.5}$ with $T_c$=45
K. The sample was grown in a BaZrO$_3$ crucible, which yields
material with purity at the 0.99995 level. The high purity allows
allows the development of very long correlation lengths
($\xi_a=148 \AA$, $\xi_b=430 \AA$, $\xi_c=58 \AA$) of Ortho II
order \cite{liang}, in which the the charge reservoir layer
consists of alternating filled and empty copper oxygen chains.
Together with YBa$_2$Cu$_4$O$_8$, it is one of two underdoped
cuprate superconductors in which doping does not introduce
disorder.  The relative lack of disorder is reflected in the low
temperature transport properties, which indicate quasiparticle
scattering times in excess of 30 ps \cite{turner}.

In our experiments, the pump and probe beams were produced by a
mode-locked Ti:Sapphire laser.  The pulses have duration 100 fs,
repetition rate 12 ns, and center wavelength 800 nm. Both pump and
probe beams were focused onto the sample with a 20 cm focal length
lens, yielding a spot size of 75 $\mu$m diameter. The reflected
probe beam was focused onto a Si photodiode detector.

A double modulation scheme was used to optimize sensitivity to
small changes in the power of the reflected probe beam. In order
to minimize the noise from the Si photodiode detector, a
photoelastic modulator was used to modulate the amplitude of the
pump beam at 100 KHz. In addition, a galvanometer-mounted mirror
varied the path-length difference between pump and probe beams at
a frequency of 40 Hz. This yields a rapid scan of the time delay
between pump and probe, which helps to suppress noise due to 1/f
fluctuations of the probe beam power. In order to demodulate and
extract $\Delta R$ as a function of time delay, the output of the
Si photodiode was sent to a lock-in amplifier for phase-sensitive
detection of the 100 KHz component. The output of the lock-in was
then sent to a digital oscilloscope whose time base was triggered
synchronously to the oscillating mirror.  With the oscilloscope in
averaging mode, a minimum detectable $\Delta R/R$ of $\approx
10^{-7}$ could be achieved after approximately 10 minutes of
accumulation time.

In this study, we focused on the decay rate of $\Delta R$
immediately following the pulsed excitation, or $\gamma(0)$. For
most of the measurements, we determined $\gamma(0)$ by fitting
$\Delta R(t)$ at small time delays by a decaying exponential.
However, special considerations arose in measurements of
$\gamma(0)$ at very low pump power. As described in succeeding
sections, $\gamma(0)$ decreases as the pump intensity is lowered.
At low temperatures, the decay rate decreases to the extent that
the change in $\Delta R$ over the 25 ps time delay range produced
by the oscillating mirror is extremely small. To measure
$\gamma(0)$ in this regime, we switched to a detection scheme that
probes directly the derivative of $\Delta R$ with respect to time
delay. The signal from the first lock-in was sent to a second
lock-in, rather than the digital oscilloscope, for phase-sensitive
detection at the frequency of the oscillating mirror. When the
oscillating time delay is less than the decay time of $\Delta R$,
the output of the second lock-in is proportional to the derivative
$d\Delta R/dt$. In this mode of data acquisition we vary the
pump-probe delay by a conventional system of a retroreflector
mounted on a translation stage.

When measured at low pump intensity, $\gamma(0)$ is remarkably
sensitive to $T$, decreasing about an order of magnitude between
15 and 10 K, for example. The extreme sensitivity to temperature
suggests that laser heating can introduce significant error in
determining the $T$-dependence of $\gamma(0)$. (By laser heating
we refer to the steady-state increase in sample temperature due to
the time-averaged laser power). For the experiments described
above, the pump intensity was lowered below the intensity of the
probe. In this regime, the probe beam is responsible for heating
the photoexcited region. At low temperature we can readily observe
that $\gamma(0)$ varies as the probe intensity, and consequently
the sample temperature, is varied. The obvious remedy of reducing
the probe power has the drawback that the $\Delta R$ signal soon
disappears below the detector noise level.

Empirically, we found that the optimal compromise between
signal-to-noise ratio and laser heating is obtained at a probe
power of approximately 1.2 mW.  At this power level $\gamma(0)$
can be measured, yet the induced temperature change is
significantly less than the bath temperature. Determining the
dependence of $\gamma(0)$ on $T$ at this power level requires that
we know the temperature of the photoexcited region of the sample.
Unfortunately it is extremely difficult to measure, with the
required accuracy, the temperature of the sample directly under
the focal spot.

Because of this difficulty, we have performed a numerical
simulation of the thermal diffusion equation to estimate the
laser-induced temperature change. We input to this simulation a
realistic model for the thermal properties of the sample and its
coupling to the bath. In the experiment, the sample is attached to
a sapphire plate using thermal grease. After mounting, the
thickness of the grease layer is determined using an optical
microscope. The input parameters to the numerical simulation are
the thicknesses of the sample and grease layer, as well as their
temperature dependent thermal conductivities
\cite{hagen,kreitman}. Figure 1 shows the temperature of the
photoexcited region as a function of the bath temperature, as
determined by finite-element analysis of the diffusion equation.
In the simulation, the laser power is 1.2 mW, focused to a spot of
diameter 75 $\mu$m. The thickness of the sample and thermal grease
were 20 $\mu$m and 5 $\mu$m, respectively. In the subsequent
analysis of the low-pump intensity data, we used the calculated
temperature, rather than the temperature of the thermal bath.

\begin{figure}
\includegraphics[width=3.25in]{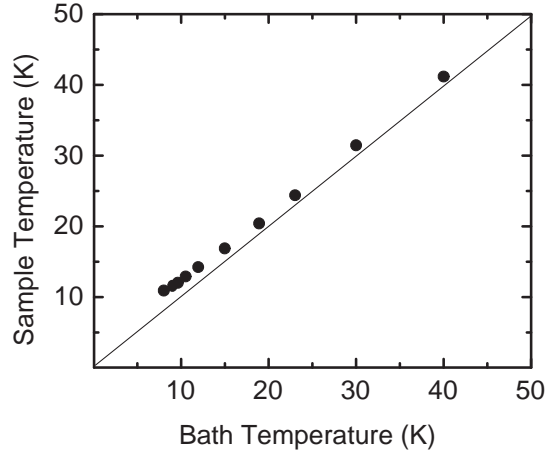}
\caption{Temperature of the photoexcited region of the sample as a
function of the bath temperature, as determined by numerical
simulation.} \label{fig:Fifth}
\end{figure}

\section{\label{sec:level1}Background: Rothwarf-Taylor equations and the phonon-bottleneck}

In this section we introduce the Rothwarf-Taylor \cite{rt}
equations, which provide a successful phenomenological framework
for understanding nonequilibrium dynamics in superconductors
\cite{gray}. The RT equations are a pair of rate equations that
describe a system of superconducting quasiparticles coupled to
phonons. In the RT phenomenology the excited state is
characterized by number densities rather than nonequilibrium
energy distribution functions. This drastic simplification is
justified in s-wave superconductors because quasiparticles rapidly
thermalize to a narrow range of energy just above the gap. In
d-wave superconductors the time evolution of the quasiparticle
distribution function may be more complicated. In spite of this,
we have found that the RT equations provide an excellent
description of the dynamics of YBCO Ortho II after pulsed
photoexcitation. In the final section we comment on the underlying
reasons for the applicability of the RT approach.

In this work we write the RT equations in the following form,

\begin{equation} \dot{n}=I_{qp}+2N\gamma_{pc}-\beta n^2
\label{eq:first} \end{equation}
\begin{equation} \dot{N}=I_{ph}+\beta n^2/2-\gamma_{pc}N-(N-N_{eq})\gamma_{esc}
\label{eq:second}
\end{equation}
where $n$, and $N$, are the number densities of gap energy
quasiparticles and phonons, respectively. The right hand side of
Eq. 1 expresses the difference between the rates of quasiparticle
creation and annihilation. $I_{qp}$ is the external generation
rate and 2$N\gamma_{pc}$ is the rate of pair creation \textit{via}
annihilation of gap energy phonons. The quasiparticle annihilation
rate $\beta n^2$ varies quadratically with density because
recombination is a two-particle scattering event. In Eq. 2 the
time rate of change of the phonon density is given by the external
phonon creation rate and the same recombination and pair creation
terms (with opposite signs) as in Eq. 1.  The parameter
$\gamma_{esc}$, which appears in the last term in Eq. 2, is the
rate at which a gap energy phonon decays into below-gap energy
phonons that cannot regenerate a quasiparticle pair. Typically,
$\gamma_{esc}\ll\gamma_{pc}$, in which case a gap energy phonon is
far more likely to regenerate a quasiparticle pair than to decay
into the bath.  Although the escape process is relatively slow, it
is essential for the ultimate return of the coupled system to
equilibrium.

In the present paper, we are concerned with the evolution of the
system following pulsed excitation, in which case there are two
characteristic time regimes.  For delay times $t$ such that
$t\ll\gamma_{pc}^{-1}$, the quasiparticles and phonons have not
yet reached quasiequilibrium. In this regime the quasiparticle
population decays at the density dependent rate $\beta n$.  For
$t\gg\gamma_{pc}^{-1}$ the exchange of energy between the phonons
and quasiparticles has brought the two populations into
quasiequilibrium. In this regime $N$ and $n$ both decay at the
much slower rate $\gamma_{esc}$. The decay of the quasiparticle
population has become slow and independent of density, despite the
fact that the rate of quasiparticle scattering may be rapid and
density dependent. The limit on the decay of the nonequilibrium
quasiparticle density imposed by the quasiequilibrium with the
phonons is termed the phonon bottleneck.

\section{\label{sec:level1}Experimental results}

\subsection{\label{sec:level2}Intensity dependence at low temperature}

The decay rate of the photoinduced reflectivity in YBCO Ortho II
depends strongly on both the temperature, $T$, and the pump
intensity, $I$.  Fig. 1 illustrates the decrease of the decay rate
with decreasing $I$ at a fixed $T$=9 K. Each curve, measured using
a different $I$ in the range from 2 to 25 mW, has been normalized
to the same value at $t=0$ to illustrate the variation in decay
rate. Although the decay rate continues to decrease as $I$ is
lowered below 2 mW, these curves are not shown because their noise
fluctuations obscure the higher intensity curves.  We discuss the
decay rate at very low pump intensity in more detail in Section
IV.C.

\begin{figure}
\includegraphics[width=3.25in]{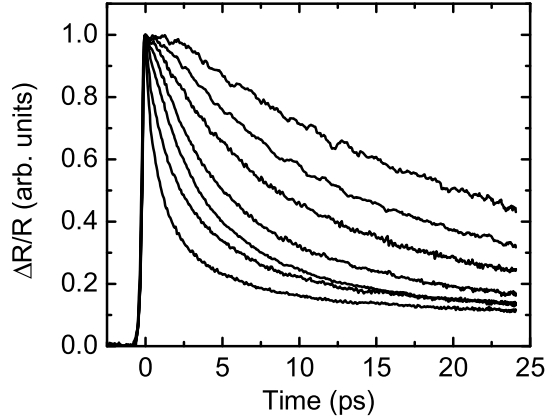}
\caption{$\Delta R/R$ at T=9 K as a function of time following
pulse excitation, for several pump intensities. All curves were
normalized to the same value at delay time zero to illustrate the
variation in initial decay rate. The most rapid decay is seen at
the highest pump intensity, corresponding to 25 mW of laser power
focused to spot of diameter $75 \mu m$. The decay becomes
systematically slower as the intensity is reduced by factors 0.69,
0.44, 0.24, 0.15, 0.10, and 0.06.} \label{fig:First}
\end{figure}

The variation in decay rate with excitation density observed in
YBCO Ortho II is highly unusual.  Analogous experiments on s-wave
superconductors have consistently found that the decay rate of the
nonequilibrium state is independent of the quasiparticle density
\cite{gray}. The lack of density dependence is understood as a
manifestation of the phonon bottleneck effect. As discussed in the
previous section, the bottleneck sets in when $t$ reaches
approximately $\gamma_{pc}^{-1}$. If there is no significant
recombination on this time scale, that is if $\beta n \ll
\gamma_{pc}$, then only density-independent bottleneck dynamics
can be seen.

Within this picture, the absence of a bottleneck in YBCO Ortho II
indicates that the hierarchy of pair creation and recombination
rates is reversed as compared with conventional superconductors.
(In Section V.D we discuss why the hierarchy of rates may be
reversed in cuprate superconductors). If $\beta n \gg
\gamma_{pc}$, then in the time window $t<\gamma_{pc}^{-1}$
quasiparticle recombination proceeds in the absence of the reverse
process of pair creation. Measuring the decay of the quasiparticle
density in this time window provides direct information about the
rate of quasiparticle-quasiparticle scattering.

To test whether our measurements of YBCO Ortho II are in a
pre-bottleneck regime, we have analyzed the time-dependence of the
decay of $\Delta R$. Fig. 3 is a double log plot of $\Delta R$ as
a function of time delay, for several pump intensities.  At the
highest excitation density $\Delta R$ decays as a power law for
$t>3$ ps. With decreasing excitation density the onset of power
law decay shifts to longer times.  At the lowest intensity the
crossover time is greater than the maximum time delay in this set
of measurements, which is $\sim$25 ps.

\begin{figure}
\includegraphics[width=3.25in]{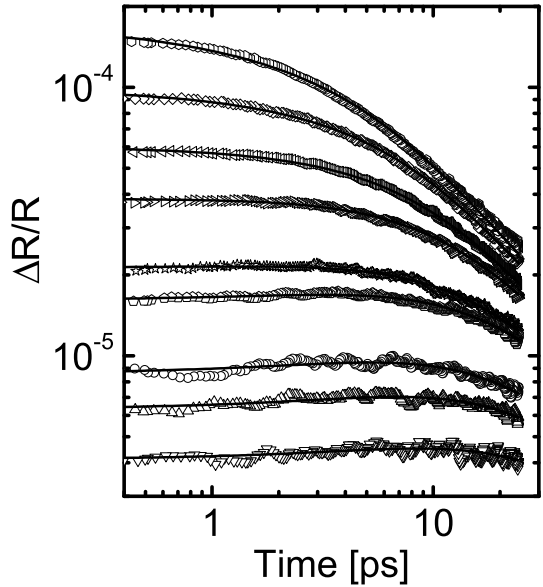}
\caption{$\Delta R/R$ \textit{vs.} time delay at T=9K plotted on a
double logarithmic scale, for pump laser powers (in mW) 0.1, 0.2,
0.4, 0.6, 1.0, 1.5, 2.4, 3.8, 6.0.} \label{fig:Second}
\end{figure}

We compare these curves with the predictions of the RT equations
in the time regime before the onset of the phonon bottleneck. This
decoupled regime corresponds to the $\gamma_{pc}\rightarrow0$
limit. In this limit the form of the RT equations is clarified if
the total quasiparticle population is written as $n_{ph}+n_{th}$,
where $n_{ph}$ and $n_{th}$ are the photoinduced and thermal
equilibrium quasiparticle density, respectively.  Similarly, we
define $N_{ph}$ and $N_{th}$ as the photoinduced and thermal
equilibrium phonon populations. Detailed balance relates the
thermal densities of phonons and quasiparticles such that,

\begin{equation} 2N_{th}\gamma_{pc}=\beta n_{th}^2.
\label{eq:third}
\end{equation}
Substituting the above relation into Eq. 1, and setting
$\gamma_{pc}=0$, yields a rate equation for $n_{ph}$ that is
independent of the phonon population,

\begin{equation}\dot{n}_{ph}=I_{qp}-\beta n_{ph}^2-2\beta n_{th}n_{ph}.
\label{eq.fifth}
\end{equation}
Eq. 4 is equivalent to the rate equation applicable to
nonequilibrium electrons and holes in semiconductors, and has been
studied extensively in connection with photoconductivity
\cite{bube}. If $n_{th}$ vanishes as $T$ approaches zero, then the
rate equation approaches $\dot{n}_{ph}=I_{ph}-\beta n_{ph}^2$. At
low $T$ the quasiparticles obey simple second order, or
bimolecular, reaction kinetics.  Integration yields the decay
after pulsed excitation,
\begin{equation}n_{ph}(t)={n_{ph}(0)\over[1+\beta n_{ph}(0)t]}.
\label{eq:third}
\end{equation}

Eq. 5 predicts that the excited population approaches $1/\beta t$
for $t\gg 1/\beta n_{ph}(0)$, regardless of the initial
quasiparticle density. However, as is clear from Fig. 3, the
measured decay curves do not obey this prediction. At long times
the decay is closer to $t^{-0.8}$ than $t^{-1}$, and curves for
different intensities do not merge to a single curve at long
times.

The discrepancies described above can be traced to a complication
omitted from the preceding analysis. Because the intensity of the
pump beam decreases exponentially with increasing depth below the
sample surface, $z$, the local quasiparticle density,
$n_{ph}(z,t)$, is spatially nonuniform. The initial local density
$n_{ph}(z,0)$, equals $n_{ph}(0,0)e^{-\alpha z}$, where $\alpha$
is the absorption coefficient at the pump wavelength. Assuming
negligible diffusion of excitations in the $z$ direction, the
local excitation density decays as,

\begin{equation}n_{ph}(z,t)={n_{ph}(z,0)\over[1+\beta n_{ph}(z,0) t]}. \label{eq:third}
\end{equation}

The probe beam, whose change in reflectivity measures the excited
quasiparticle density, also decays exponentially in the sample.
Therefore the measured reflectivity change is related to an
exponentially weighted average of the nonequilibrium quasiparticle
density \cite{fishman},

\begin{equation}
\mathcal{N}(t)\equiv\alpha^{-1}\int_0^\infty dze^{-\alpha
z}n_{ph}(z,0)/[1+\beta n_{ph}(z,0) t]. \label{eq:sixth}
\end{equation}
Assuming that the change in reflectivity is proportional to
$\mathcal{N}$ finally yields,

\begin{equation}
\Delta R(t)={2\Delta R(0)\over \gamma_0 t} [1-{ln(1+\gamma_0 t)
\over \gamma_0 t}], \label{eq:seventh}
\end{equation}where $\gamma_0 \equiv \beta n_{ph}(0,0)$.

The solid lines in Fig. 3 indicate the best fit of Eq. 8 to the
experimental data.  The fit is obtained by varying $\Delta R(0)$
and $\gamma_0$ for each curve.  There is excellent agreement
between the data and the prediction of the bimolecular rate
equation when the exponential variation of the pump and probe
intensities is taken into account.  The agreement suggests that
for $t<25$ ps, the phonon bottleneck is not yet established and
the decay rate of $\Delta R$ is a direct measure of the
quasiparticle-quasiparticle scattering rate.

From the fitting procedure we determine $\Delta R(0)$ and
$\gamma_0$ for each curve.  In Fig. 4 we plot $\gamma_0$ as a
function of $\Delta R(0)/R$.  It is evident that the
characteristic decay rate increases \textit{linearly} with the
magnitude of the initial reflectivity change. The linear
relationship indicates that the entire family of decay curves can
be described by a single bimolecular coefficient $\beta$.

\begin{figure}
\includegraphics[width=3.25in]{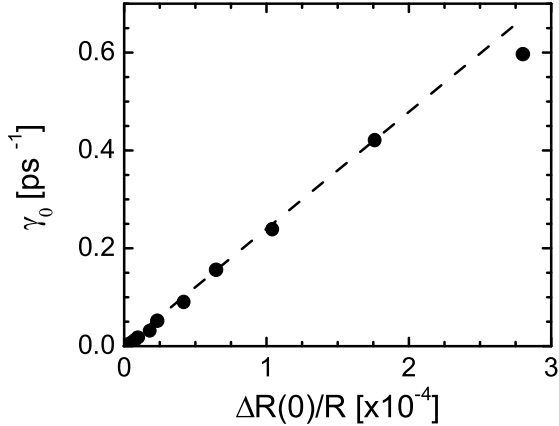}
\caption{The initial decay rate at the sample surface, $\gamma_0$,
as a function of the initial reflectivity change, $\Delta R(0)/R$.
The dashed line emphasizes the linear dependence of initial decay
rate on initial density.} \label{fig:Third}
\end{figure}

\subsection{\label{sec:level2}Temperature dependence}

In the previous section we showed that the decay of $\Delta R$ at
low temperature can be described by the bimolecular rate equation
with a single value of the scattering coefficient $\beta$.  In
this section we analyze the effect of raising the temperature on
the rate of decay of $\Delta R$. Fig. 5 shows a set of normalized
decay curves of $\Delta R$, measured in the temperature range from
5-70 K, induced by a pump power of 2.5 mW. The set of curves
illustrates clearly that the decay rate at fixed $I$ increases
rapidly with increasing $T$.

To analyze the $T$ dependence, we again consider the RT equation
in the decoupled regime (Eq. 4). At nonzero $T$ we must include
the last term on the right-hand side, which was neglected
previously. Physically, this term describes the rate of scattering
of a photoinjected quasiparticle by a thermal equilibrium one.

\begin{figure}
\includegraphics[width=3.25in]{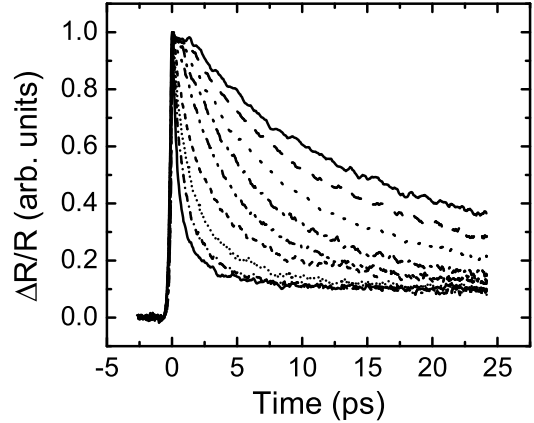}
\caption{$\Delta R/R$ induced by 2.5 mW of laser power, as a
function of time following pulse excitation, for several
temperatures in the range from 5-70 K. All curves were normalized
to the same value at delay time zero to illustrate the variation
in initial decay rate. The decay is most rapid at the highest
temperature, and becomes systematically slower as $T$ is reduced
from 70 K to 50 K, 44 K, 28 K, 22 K, 17 K, 12 K, and 5 K.}
\label{fig:Fourth}
\end{figure}
We have found that the most direct way to compare the data with
Eq. 4 is to focus on the initial decay rate, $\gamma(0)$, of the
experimentally measured transients.  Here $\gamma(0)$ is defined
as the limit of the instantaneous decay rate
$\gamma(t)\equiv-\dot{n}_{ph}/n_{ph}$ as $t$ approaches zero.
According to Eq. 4,
$\gamma(0)=\beta\left[n_{ph}(0)+2n_{th}\right]$. Assuming that the
initial reflectivity change is proportional to $n_{ph}(0)$, a plot
of $\gamma(0)$ vs. $\Delta R(0)$ should yield a straight line with
slope proportional to $\beta$ and intercept $2\beta n_{th}$.

Fig. 6 shows $\gamma(0)$ plotted as a function of $\Delta R(0)/R$
for several temperatures in the superconducting state.  The data
at the lowest temperature are essentially equivalent to the
results presented earlier.  In Fig. 4 the decay rate was
determined by fitting the entire time dependence, while  in Fig. 6
the decay rate is determined from the initial slope.  The reason
the decay rates in the two figures differ by a factor of two can
be understood from Eq. 8.  Taking the derivative of this formula
with respect to time shows that $\gamma(0)=\gamma_0/2$.

With increasing $T$ the $\gamma(0)$ vs. $\Delta R(0)$ plots shift
vertically, with little change in slope.  The fact that the slope
is nearly constant implies that $\beta$ depends weakly, if at all,
on the temperature.  The increase of the intercept with $T$
implies a rapidly increasing density of thermal equilibrium
excitations.  In the next section we examine the temperature
dependence of the intercepts, as determined from measurements
performed at very low pump and probe intensities.

\begin{figure}
\includegraphics[width=3.25in]{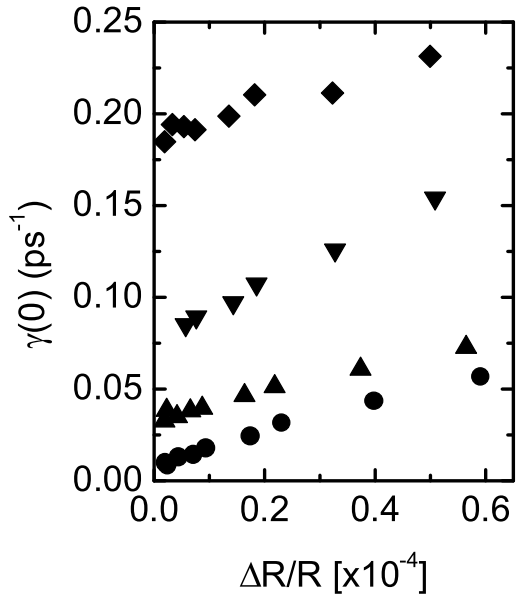}
\caption{Initial decay rate, $\gamma(0)$, as a function of initial
reflectivity change, $\Delta R(0)/R$, for T=12 K, 17 K, 22 K, and
30 K. The thermal equilibrium decay rate, obtained from
extrapolation of the data to zero $\Delta R(0)/R$ increases
rapidly with temperature.} \label{fig:Fifth}
\end{figure}

Fig. 7 presents a different perspective of the decay rate as a
function of temperature and intensity: $\gamma(0)$ as a function
of $T$ for three values of $I$. In the normal state and just below
$T_c$ the different data sets lie on the same curve, indicating
that $\gamma(0)$ is independent of $I$. As the temperature is
lowered below $T_c$ the decay rates measured with different pump
intensities begin to diverge. As the temperature tends to zero,
the decay rate crosses over to a temperature independent regime,
with the crossover temperature higher when the rates are measured
at greater pump intensity. A consistent explanation for the $T$
and $I$ dependence of $\gamma(0)$ can be found in the picture of
pairwise scattering involving both thermal equilibrium and
photoinduced quasiparticles.  The lack of intensity dependence
above $T_c$ suggests that in this regime the thermal density of
quasiparticles is far larger than the photoinduced density.  The
strong temperature dependence above $T_c$ suggests that $n_{th}$
decreases with decreasing temperature in the normal state,
possibly due to the opening of the pseudogap. The onset of
intensity dependence at $T_c$ may indicate that $n_{th}$ has
become comparable to $n_{ph}$.  Alternatively, the sudden change
in decay kinetics may be related to the onset of the coherence in
the antinodal quasiparticle self-energy.  Finally, the decay rate
measured at fixed pump intensity crosses over to $T$-independence
when $n_{th}$ becomes much smaller than $n_{ph}(T)$. The overall
behavior of $\gamma(0)$ suggests that the decay rate vanishes in
the limit that both $I$ and $T$ go to zero.

\begin{figure}
\includegraphics[width=3.25in]{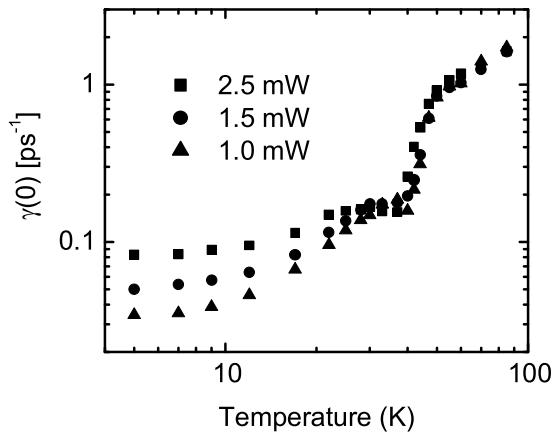}
\caption{Initial decay rate, $\gamma(0)$, as a function of $T$ for
three values of pump power.} \label{fig:Fifth}
\end{figure}

\subsection{\label{sec:level2}Low intensity regime}

In this section we focus on measurements of $\gamma(0)$ performed
at the lowest laser intensities that are accessible
experimentally. The motivation for studying the low-intensity
regime is to measure, \textit{via} a nonequilibrium experiment,
the recombination lifetime of quasiparticles in equilibrium. In
order to probe equilibrium properties, the density perturbation
introduced by the laser must be small, such that $n_{ph} \ll
n_{th}$.  According to Eq. 4, $\gamma(0)$ approaches $2\beta
n_{th}$ in this limit, which is exactly twice the thermal
equilibrium recombination rate, $\gamma_{th}$.

Measuring $\gamma_{th}$ accurately becomes increasingly difficult
at low temperature because of the requirement that $n_{ph}\ll
n_{th}$. As $n_{th}$ decreases rapidly with decreasing $T$, the
pump intensity must be lowered accordingly. Eventually $\Delta R$
reaches the noise floor of the experiment, which sets a limit on
the smallest $\gamma_{th}$ that can be measured. In principle, the
signal size could be increased by raising the intensity of the
probe beam. However, this inevitably leads to an increase in the
average temperature of the photoexcited region compared with the
bath temperature. This problem becomes increasingly severe at low
temperature, where the thermal conductivity of the sample is
small. For the data to be presented, the probe intensity was
maintained at 1.2 mW, a value which provides the best compromise
between sensitivity and heating.

At each $T$, we obtain 2$\gamma_{th}$ from the extrapolation to
zero of a linear fit to $\gamma(0)$ vs. $\Delta R$.  The
temperature of the photoexcited region is determined through the
numerical analysis described in Section II. Fig. 8 is a double
logarithmic plot of 2$\gamma_{th}$ as a function of $T$, showing
results for both YBCO Ortho II and a thin film sample of
Bi$_2$Sr$_2$CaCu$_2$O$_8$ (BSCCO) \cite{segrethesis}. At the upper
limit of the $T$ range, $\gamma_{th}$ is nearly identical in the
two materials. However, as mentioned previously, the dependence of
$\gamma_{th}$ on $T$ becomes very strong in YBCO Ortho II at low
temperature. If the functional form was assumed to be a power law,
then $\gamma_{th}$ would approach a $T^8$ dependence.

We believe it to be more reasonable to assume that $\gamma_{th}$
in YBCO Ortho II decreases exponentially at low temperature. To
analyze this exponential dependence, we compare the data with the
formula,

\begin{figure}
\includegraphics[width=3.25in]{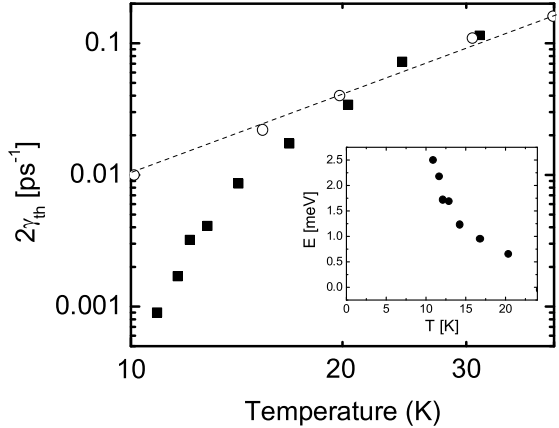}
\caption{Twice the thermal equilibrium recombination rate, as
determined by the extrapolation of $\gamma(0)$ to zero laser
power, as a function of temperature. Solid squares are results for
the YBCO Ortho II sample, open circles show results obtained on a
thin film sample of BSCCO with $T_c$= 85 K. The dashed line
indicates a temperature dependence proportional to $T^2$. The
inset shows the activation energy of the recombination rate in
YBCO Ortho II as determined from a fit to Eq. 9.}
\label{fig:Fifth}
\end{figure}

\begin{equation}
2\gamma_{th}(T)={k_B T \over \hbar}{T \over
T_0}exp\left[-\Delta_{th}(T)/k_B T\right]. \label{eq:seventh}
\end{equation}
We introduce a prefactor proportional to $T^2$ for two reasons,
one theoretical and the other experimental. The theoretical reason
is that $n_{th}(T)$ is predicted to vary as $T^2$ for a d-wave
superconductor.  The experimental reason is that in some samples
of BSCCO $\gamma_{th}$ is proportional to $T^2$ over the
temperature range from 10 K to 40 K \cite{segrethesis}. (The
dashed line in Fig. 8 indicates a $T^2$ dependence). Fitting Eq. 9
to the YBCO Ortho II data yields $T_0=1100$ K and a temperature
dependent activation energy $\Delta_{th}$ that is plotted in the
inset. The origin of the exponential cutoff in $\gamma_{th}$ that
appears in YBCO Ortho II and not in BSCCO is not understood at
present. However, we note that analogous behavior is observed in
the scattering rates observed by ac conductivity experiments. In
the YBCO system, the mean free path increases exponentially with
temperature towards a limit of several microns set by elastic
processes \cite{turner}. In BSCCO the $T$ dependence is much
weaker and the limiting value is only a few hundred $\AA$
\cite{corson}. It is possible that momentum conservation imposes
constraints on the rate of quasiparticle-quasiparticle scattering
in the cleaner YBCO system. These constraints may dictate that
only Umklapp processes are allowed at low $T$, leading to an
exponential $T$ dependence \cite{walker,howell}.

\section{\label{sec:level1} Discussion}

In analyzing the experimental data presented above, we begin (in
Section V.A) with the most basic question: what type of excitation
is probed by transient reflectivity experiments performed on the
cuprate superconductors? We argue that these excitations are
quasiparticles that occupy states close to the antinodal regions
of the Brillouin zone. In Sections V.B and V.C we turn to a
quantitative analysis of the decay rate of the photoexcited state.
Section V.B focuses on our observation that the lifetime of a
photoinjected quasiparticle diverges in the limit that $T$ and $I$
both tend to zero. We show theoretically that the constraints of
momentum and energy conservation prevent thermalization of
antinodal quasiparticles toward the nodes.  In Section V.C we
estimate the magnitude of the recombination coefficient $\beta$
from the experimental data, and discuss its theoretical
interpretation. Finally, in Section V.D, we comment on possible
reasons for the absence of a phonon bottleneck in the decay of the
photoexcited state.

\subsection{\label{sec:level2}What types of excitations are probed?}

As a first guess, one might imagine that the photoinduced
quasiparticles occupy states near the nodes, as these are the
lowest energy excitations. This intuition is consistent with the
generally accepted picture of the quasiparticle distribution
function in the nonequilibrium state. In this picture the
nonequilibrium quasiparticles rapidly adopt a Fermi-Dirac
distribution, but with a chemical potential, and temperature,
$\mu^\ast$, and $T^\ast$, respectively, that are larger than their
values in thermal equilibrium \cite{parker,owen}.  In a d-wave
superconductor this distribution describes a population of
quasiparticles that is dominated by states near the nodes ('nodal'
quasiparticles). The occupation of states with energy greater than
max($\mu^\ast, T^\ast$) is exponentially small.

On the other hand, the experimental evidence suggests that the
photoexcited state is not a degenerate gas of nodal
quasiparticles. Three observations, in particular, lead us to this
conclusion:

(1) \textit{The diffusion coefficient of photoinjected
quasiparticles is at least two orders of magnitude smaller than
the diffusion coefficient of nodal quasiparticles} \cite{gedik}.
The diffusive propagation of nonequilibrium quasiparticles in YBCO
Ortho II was recently measured by the transient grating technique.
Values for the diffusion coefficient of 20 cm$^2$/s and 24
cm$^2$/s were determined for motion along the \textbf{a} and
\textbf{b} crystalline axes, respectively. This may be compared
with a nodal quasiparticle diffusion coefficient of $\approx$ 6000
cm$^2$/s as determined from microwave spectroscopy on the same
sample \cite{turner}.

(2) \textit{The reduction in the condensate spectral weight
depends linearly on the pump intensity.} Information about the
photoinduced reduction of condensate spectral weight comes from
visible pump-terahertz probe experiments \cite{averitt,carnahan}.
At frequencies $\sim$1 THz, the optical conductivity
$\sigma(\omega)$ is dominated by its imaginary part, $\sigma_2$,
which is proportional to the condensate density. Following
photoexcitation at 1.5 eV, $\sigma_2$ drops rapidly as the
condensate density is diminished. At intensities below the
saturation level, the loss of condensate density is a linear
function of the energy deposited in the photoexcited volume.

The linear reduction of condensate density with pump energy is
inconsistent with the physics of nodal quasiparticles. This point
can be illustrated by considering the temperature dependence of
the superfluid density, $\rho_s$.  Nodal quasiparticles are
responsible for the linear in $T$ decrease of $\rho_s$ observed in
clean cuprate superconductors \cite{hardy}. However, the energy
stored in the nodal quasiparticle gas increases not as $T$, but as
$T^3$. Therefore the reduction in $\rho_s$ due to nodal
quasiparticles varies as the one-third power of the total energy
in the quasiparticle gas. We would expect such a strongly
sublinear relationship for a degenerate gas of nonequilibrium
nodal quasiparticles as well.

(3) \textit{Far-IR measurements of the photoexcited state show
that the optical spectral weight that is removed from the
condensate by photoexcitation shifts to energies on the order of
50-100 meV, or several times the value of the maximum gap}
\cite{kaindl}. As discussed above, thermal excitation of nodal
quasiparticles does indeed remove spectral weight from the
condensate. However, the spectral weight shifts to a very narrow
($<$30$\mu$eV) Drude peak centered on $\omega=0$. In contrast, the
spectral weight removed from the condensate by photoexcitation
shifts to frequencies that are $\approx$1000 times higher than the
width of the nodal quasiparticle Drude peak.  This is further
evidence against a degenerate gas of nodal quasiparticles.

On the basis of the preceding arguments, we conclude that nodal
quasiparticles do not dominate the population of quasiparticles
created by photoexcitation.  In other words, the quasiparticle
distribution in the photoexcited state cannot be described as a
Fermi-Dirac function with effective parameters $\mu^\ast$ and
$T^\ast$. The experiments are more consistent with a distibution
function that is peaked at an energy above the chemical potential.
Such a distribution can arise if it is not possible to scatter
into the the nodal states during the lifetime of the photoexcited
state.  In Section 5.2 we show that the constraints of energy and
momentum conservation severely restrict the rate at which hot
quasiparticles scatter into states near the nodes.

\subsection{\label{sec:level2}Stability of an isolated photoexcited particle}

In section V.A, we argued that the quasiparticle distribution
function in the photoexcited state peaks at an energy above the
chemical potential and therefore is not of the Fermi-Dirac form.
Such a nonequilibrium distribution can arise if quasiparticles
cannot scatter into nodal states during the lifetime of the
photoexcited state. However, we have also shown that this lifetime
can be very long, reaching $\sim$ 0.6 ns at the lowest temperature
and intensity accessible experimentally. By comparison a 30 meV
excitation in a Fermi liquid (with Fermi energy $\sim$ 1 eV) would
decay into electron-hole pairs in $\sim$3 ps. In the following we
describe how the relative stability of a quasiparticle in a d-wave
superconductor can arise from the severely restricted phase space
for decay.

In a Fermi liquid, quasiparticles readily decay into particle-hole
pairs because energy and momentum can be conserved in the process.
However, the kinematic contraints are much more severe in the case
of a $d$-wave superconductor. Consider the momentum-resolved
particle-hole excitation spectrum in the two cases.  In the Fermi
liquid the particle-hole spectrum forms a broad continuum that
extends from zero wavevector to twice $k_F$, even as the pair
energy tends to zero.

The d-wave superconductor differs from the Fermi liquid in that
the Fermi contour shrinks to four nodal points. As a consequence,
the pair excitation spectrum is much more localized in momentum
space. Fig. 9 illustrates the momentum-resolved pair excitation
spectrum of a d-wave superconductor.  Each plot is a color-scale
depiction of the density of two-particle excitations of a given
energy in the first Brillouin zone. The spectra were calculated
using a parameterization of the quasiparticle dispersion obtained
by Norman \cite{norman} from a fit to ARPES data.  At very low
energies, the pair spectrum is localized near nine discrete
wavevectors that represent the creation of particles at the nodes.
As the energy of the pair increases, the locus of possible
wavevectors spreads out. However, for energies that are less than
$\Delta_0$, the allowed momenta remain clustered near the nine
zero-energy wavevectors.

In order for a single quasiparticle to scatter and emit a pair,
its change in energy and momentum, $\Delta\epsilon$ and $\Delta
k$, must match that of the pair, $\epsilon_{pair}$ and $k_{pair}$.
The constraints imposed by this condition are illustrated in the
diagrams shown in Fig. 10, which describe the possible decay
events of a hot quasiparticle. For this example we have chosen a
quasiparticle with energy 40 meV and momentum on the Fermi contour
of the normal state. Each diagram corresponds to a fixed energy
transfer in the scattering event. The color scale indicates the
location of allowed values of $\Delta k$ and $k_{pair}$ at the
transfer energy. A scattering event is kinematically allowed only
at momentum transfers where the two sets of wavevectors overlap.

\begin{figure*}
\includegraphics[width=6.75in]{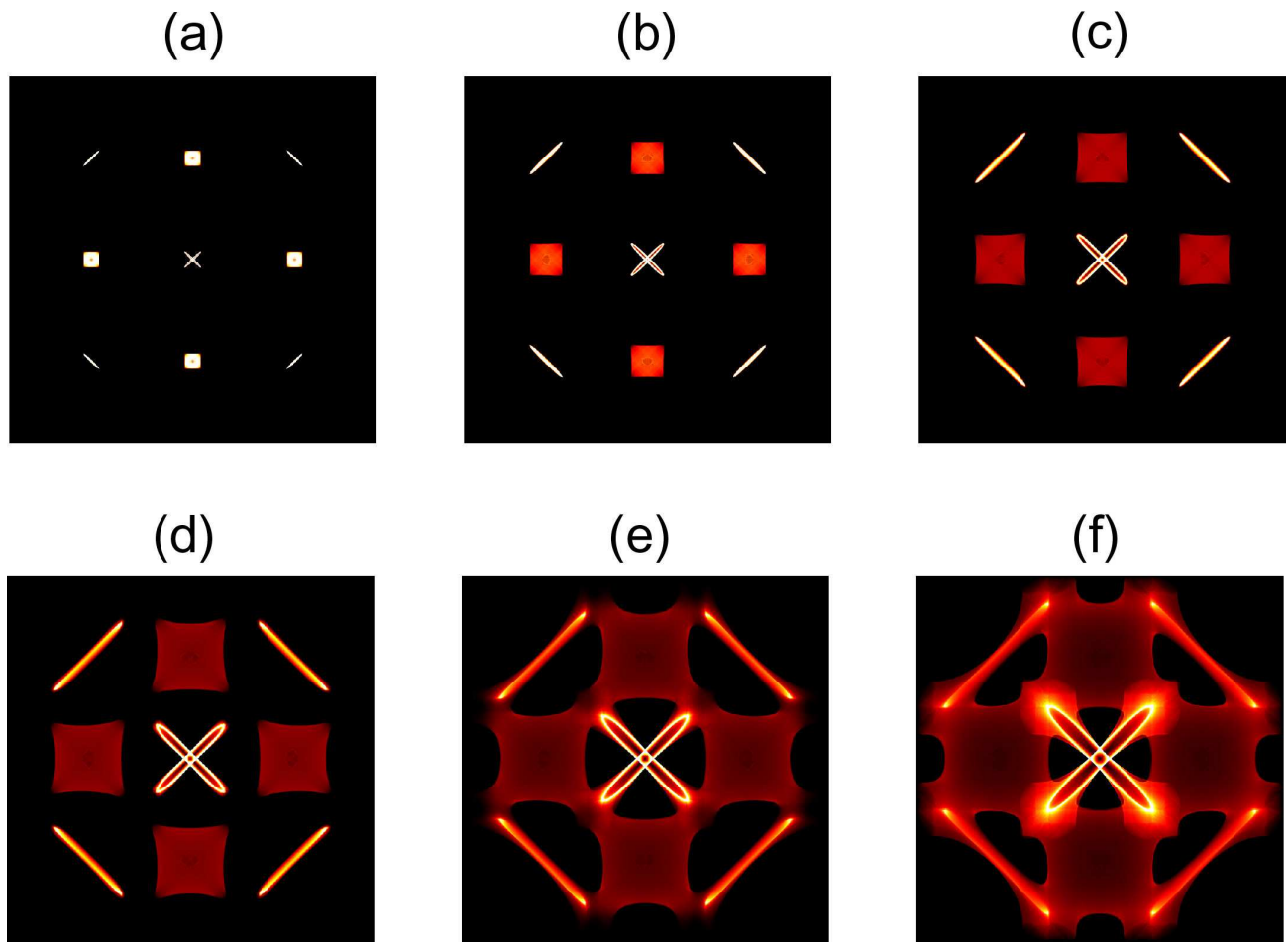}
\caption{Color scale depiction of quasiparticle pair momenta in
the first Brillouin zone for different values of the pair energy;
(a) 10 meV, (b) 20 meV, (c) 30 meV, (d) 40 meV, (e) 50 meV, and
(f) 60 meV.} \label{fig:Fifth}
\end{figure*}

Fig. 10 illustrates how kinematics prevent the inelastic
scattering of the hot quasiparticle.  For example, consider the
scattering events with the largest possible energy change, 40 meV.
To give up all its energy the hot quasiparticle must scatter to
one of the nodes, thus there are only four allowed values of
$\Delta k$. (The four points are not visible in Fig. 10(a)). One
of these four wavevectors just touches the perimeter of the
elliptical region of pair wavevectors, which indicates an allowed
scattering event.  In this scattering event the hot quasiparticle
scatters to the node, with the simultaneous creation of a
quasihole at the node and a quasiparticle at the energy and
momentum of the hot quasiparticle. This exchange process, although
kinematically allowed, does not lead to a change in the particle
distribution function.

\begin{figure*}
\includegraphics[width=6.75in]{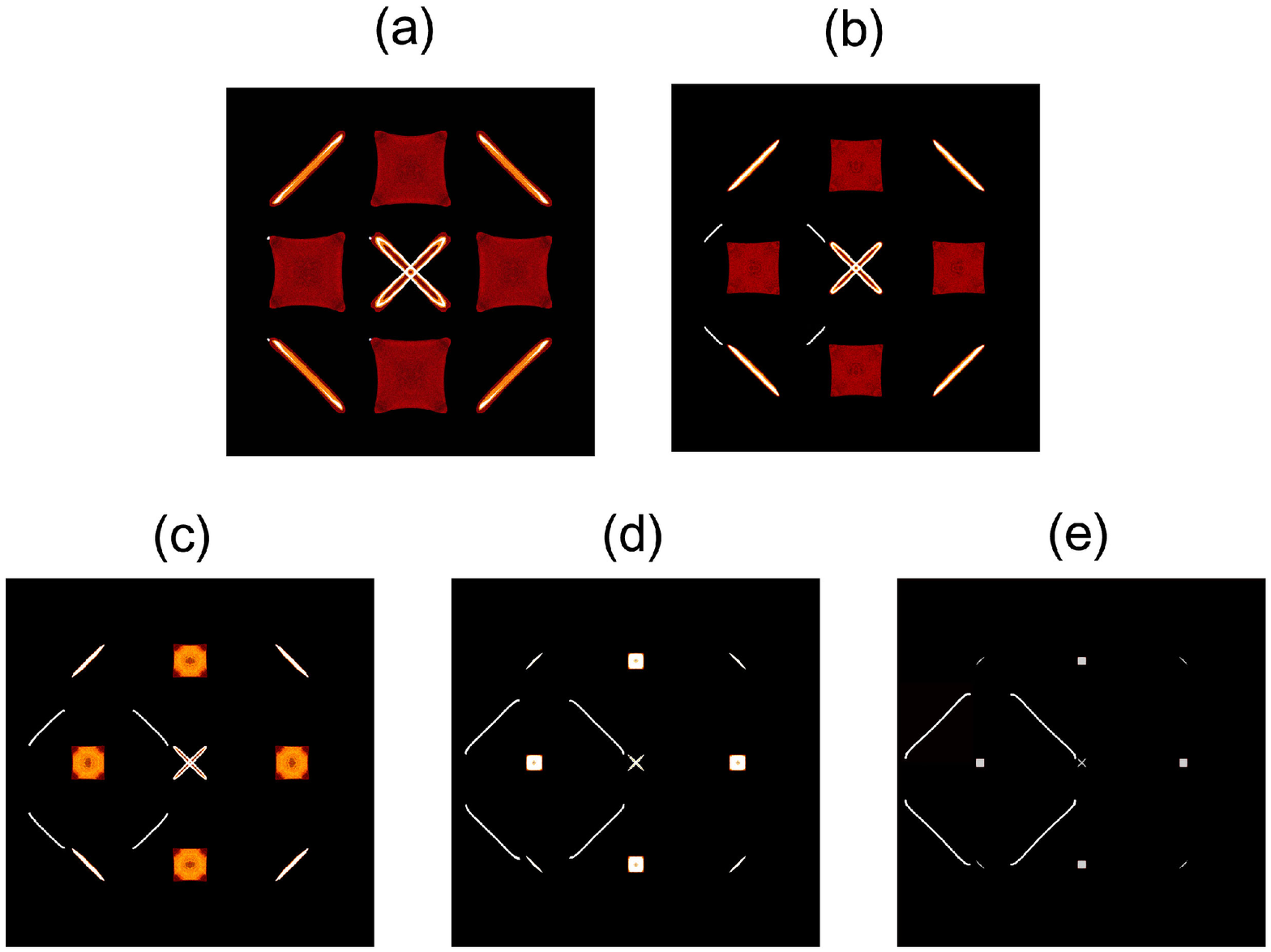}
\caption{Depiction of the lack of overlap between the scattering
momentum $\Delta k$ and the pair momentum $k_{pair}$. For this
illustration the hot quasiparticle starts with energy 40 meV and
momentum on the Fermi contour of the normal state. The arc
segments in each square show the possible values of $\Delta k$ for
several values of the scattering energy $\Delta\epsilon$: (a) 40
meV; (b) 30 meV; (c) 20 meV; (d) 10 meV; (e) 5 meV.  The color
scale plot in each of the squares shows the allowed values of
$k_{pair}$ at the same energy. As the energy transfer decreases
and the arc segments expand, the regions of $k_{pair}$ shrink
faster.  The lack of overlap is a graphical demonstration of how
momentum and energy conservation prevent thermalization of a
quasiparticle near the antinodal region of the Brillouin zone.}
\label{fig:Fifth}
\end{figure*}

If the hot quasiparticle does not lose all of its energy, there
are more choices for the final momentum. The locus of allowed
$\Delta k$ expands, becoming visible as arcs in panels (b) through
(e) of Fig. 10. However, as the arcs expand the locus of pair
creation wavevectors shrinks. Fig. 10 shows that the region of
allowed $k_{pair}$ shrinks faster than the expansion of allowed
$\Delta k$, so that the two regions fail to overlap at any energy
transfer less than the energy of the hot quasiparticle. The
absence of overlap demonstrates that the thermalization of the hot
quasiparticle is forbidden because of kinematic constraints.
Recently, Howell et al. \cite{howell} reached a similar conclusion
based on an analytical, rather than numerical, approach.

The kinematic constraints in this example are specific to
quasiparticles near the normal state Fermi contour. They are a
consequence of the rapid decrease of quasiparticle velocity with
increasing distance from the node. For this reason, we anticipate
that not all quasiparticles of a given energy will be stable. The
contours of constant quasiparticle energy are distorted ellipses
centered on each of the nodes. Quasiparticles near the major axis
vertex of the ellipse will be stable, as we have seen. However,
the velocity for particles near the minor axis vertex is
comparable the nodal quasiparticle velocity. For these
quasiparticles there are no kinematic constraints to prevent rapid
thermalization. Thus, we anticipate that after a short time
following pulsed injection only nonequilibrium particles near the
major axis vertex will survive.

\subsection{\label{sec:level2}Bimolecular recombination}
\subsubsection{\label{sec:level3}Estimating $\beta$}

In the previous section we have seen that kinematic constraints
can stabilize an isolated antinodal quasiparticle.  However, our
experiments show that a \textit{pair} of quasiparticles can
convert to a state which no longer contributes to $\Delta R$. We
can be confident that this conversion requires two particles, and
not more, because the decay rate increases linearly with density.
The coefficient that relates the decay rate to the density,
$\beta$, is directly related to the cross-section for inelastic
scattering of two quasiparticles.  Thus the magnitude of $\beta$
is of fundamental importance as a measure of the coupling of
quasiparticles to some other excitation of the interacting
electronic system.

We turn next to estimating $\beta$ from the measured dependence of
the decay rate on density.  Experimentally, we measure both the
nonequilibrium decay rate $\beta n_{ph}$ and the thermal
equilibrium rate $\beta n_{th}$.  To extract $\beta$ from these
measurements, we need to know either $n_{ph}$ or $n_{th}$.
Unfortunately, neither quantity is directly determined from the
data, and we need additional assumptions to estimate these
quantities.  To estimate $n_{ph}$ we use energy conservation to
convert from laser energy density to quasiparticle density.  In
the linear regime of $\gamma$ vs. $I$ a recombination rate of
$10^{11}$ s$^{-1}$ results from a laser energy density of 0.1
J-cm$^{-3}$.   If photon energy is converted entirely to 40 meV
quasiparticles, then their initial density is $\approx 10^{12}$
cm$^{-2}$, and therefore $\beta \approx 0.1$ cm$^2$/s.

We can also determine $\beta$ from the thermal equilibrium
recombination rate as well, if we know $n_{th}(T)$. For a d-wave
superconductor,

\begin{equation}n_{th}(T)={\pi v_F \over 6v_2}\left({k_BT \over \hbar v_F}\right)^2, \label{eq:third}
\end{equation} where $v_F$ and $v_2$ are the nodal quasiparticle
velocities perpendicular and parallel to the Fermi contour,
respectively \cite{mchiao}. Based on the RT equations, we would
expect the thermal equilibrium decay rate to be proportional to
$n_{th}$ and thus vary with temperature as $T^2$. Instead, we find
experimentally that in YBCO Ortho II the limiting behavior of
$\gamma_{th}$ is exponential rather than power law. The unexpected
exponential cutoff of the recombination rate may be related to a
similar exponential cutoff seen in the momentum scattering rate as
measured through microwave spectroscopy \cite{hosseini}.

Despite the exponential cut-off of the recombination rate at low
$T$, we can attempt to estimate $\beta$ from the thermal
recombination rate at higher $T$.  Above $\approx$ 20 K,
$\gamma_{th}(T)$ is not inconsistent with a quadratic dependence
on temperature. In that range we found that
$2\gamma_{th}(T)\approx(k_BT/\hbar)T/T_0$, with $T_0=$1100 K.
Assuming that $\gamma_{th}=\beta n_{th}$, and using Eq. 10 to
estimate $n_{th}(T)$ (with $v_F/v_2=10$) we find $\beta=0.26$
cm$^2$/s. Considering the approximations and assumptions involved,
this value is quite close to the one determined from the
nonequilibrium recombination rate.

\subsubsection{\label{sec:level2}Theoretical approaches for
$\beta$}

Within the context of BCS theory, the process of recombination is
the annihiliation of a quasiparticle pair with the simultaneous
emission of a gap energy phonon.  The theory of the recombination
rate of BCS quasiparticles was developed by Kaplan et al.
\cite{kaplan}, whose calculations assumed an isotropic s-wave gap
and the breaking of momentum conservation through strong elastic
scattering ('dirty limit').  Their result for the recombination
rate, expressed through our definition of recombination
coefficient, is $\beta=4\pi\alpha^2(2\Delta)F(2\Delta)/\hbar
N(0)$, where $\alpha$, $F$, and $N$, are the electron-phonon
coupling strength, density of phonon states, and density of
electron states, respectively.  While it is very useful to have
this explicit result to compare with, we must bear in mind that
the cuprate superconductors do not satisfy the assumptions upon
which this calculation is based.  The cuprates are 2D rather than
3D, d-wave rather than s-wave, and are in the clean, rather than
the dirty, limit of elastic scattering.

We are not aware of a calculation of the phonon-mediated
recombination rate applicable to the cuprate superconductors.
However, in anticipation of such a calculation, we define a
dimensionless constant $C\equiv\hbar\beta N(0)$, that parametrizes
the electron-phonon coupling strength in the clean limit.  If in
addition we express the 2D electronic density of states through an
effective mass $m^\ast$, then we obtain the simple formula
$\beta=C\hbar/m^\ast$. Assuming an effective mass that is three
times the free electron mass, $\hbar/m^\ast=0.3$ cm$^2$/s, so that
$\beta=0.1$ cm$^2$/s corresponds to $C\approx 1/3$. It is striking
that $C$ is of order unity. This suggests that, while we do not
know if phonon emission is indeed the mechanism for quasiparticle
recombination, we have identified $\hbar/m^\ast$ as the 'natural
unit' for $\beta$.

The body of experimental results on the cuprate superconductors
have led many researchers to conclude that quasiparticle
scattering is determined by electron-electron, rather than
electron-phonon interactions.  The BCS theory of recombination was
adapted by Quinlan et al. \cite{scalapino} to the case where
quasiparticle-pair energy is converted to antiferromagnetic spin
fluctuations, rather than phonons.  As we explain below, the
results of this calculation cannot be compared directly with
$\beta$.  The recombination rate at low $T$ is predicted to vary
as $T^3$. Therefore $\gamma_{th}$ is not proportional to $n_{th}$,
which, as we have seen, varies as $T^2$.  Expressed in the
language of this paper, $\beta$ vanishes in the limit that $T$
goes to zero, in contrast to the $T$-independent value observed
experimentally.

The prediction that $\gamma \propto T^3$ while $n_{th} \propto
T^2$ is a reflection of the underlying Fermi liquid theory (FLT).
In FLT the normal state scattering rate varies as $T^2$, while the
thermal quasiparticle density varies as $T$.   The corresponding
quantities vary as one higher power of $T$ for the d-wave
superconductor  because its density of states is linear in energy,
rather than contant.  Thus it is a general consequence of the
phase space of Fermi liquids that the quasiparticle scattering
rate is not simply proportional the quasiparticle density.

It is interesting, and perhaps relevant, to recall here the
celebrated fact that the high-$T_c$ cuprates violate this central
prediction of FLT.  That is to say, the normal state scattering
rate varies linearly, rather than quadratically, with the
temperature.  In this marginal Fermi liquid (MFL) \cite{varma},
the scattering rate $\sim 2k_BT/\hbar$ \textit{is} proportional to
the quasiparticle density $N(0)k_BT$.  The coefficient of
proportionality, or $\beta_{MFL}$, is simply $\hbar/\pi m^\ast$.
Evaluating this expression with $m^\ast =3 m_0$ yields
$\beta_{MFL}$=0.1 cm$^2$/s, which is remarkably close to the
experimental value. This agreement suggests that there may be a
close connection between the quasiparticle scattering in the
normal state and the antinodal quasiparticle recombination in the
superconducting state.

\subsection{\label{sec:level2}Absence of phonon bottleneck}

In the previous section we described two mechanisms for
quasiparticle pair recombination. In the more familiar one, the
pair is destroyed with simultaneous creation of a 2$\Delta$
phonon. In the second mechanism the pair is converted to an
electronic excitation, such as an antiferromagnetic fluctuation.
Both mechanisms can provide an explanation for the absence of a
bottleneck. In the second mechanism, the quasiparticle pair decays
to other particle-hole excitations.  If the number of particles
generated in the pair-destruction process is two or more, then it
is highly unlikely for the reverse process to occur. The low rate
for the reverse process leads to the absence of a bottleneck.

On the other hand, phonon-mediated recombination is inherently
susceptible to a bottleneck because only one particle (the phonon)
is created when the pair is destroyed. Following pulsed injection,
the bottleneck sets in when the pair destruction and creation
rates become equal, or $\beta n_{ss}^2=2\gamma_{pc}N_{ss}$, where
$n_{ss}$ and $N_{ss}$ are the quasiequilibrium steady-state
densities of quasiparticles and $2\Delta$ phonons, respectively.
The size of $n_{ss}$ relative to the initial density $n_0$ is a
measure of the strength of the bottleneck. This ratio can be
calculated using conservation of energy, which implies that
$2n_{ss}+N_{ss}=2n_0$. The result is

\begin{equation} {n_{ss}\over n_0}={\lambda \over
2}(\sqrt{1+{4\over\lambda}}-1), \label{eq:third} \end{equation}
where $\lambda\equiv2\gamma_{pc}/\beta n_0$.  The strong
bottleneck regime corresponds to $\lambda\gg1$, in which case the
steady state density nearly equals the initial density,
$n_{ss}/n_0\approx1-1/\lambda$. On the other hand, if
$\lambda\ll1$ the bottleneck is practically nonexistent and the
steady state density is very small,
$n_{ss}/n_0\approx\lambda^{1/2}$.

To determine whether the cuprates are in the strong or weak
bottleneck regime, we must estimate the magnitude of $\lambda$. To
do so we first recognize that $\lambda$ must exceed a
$\lambda_{min}$, obtained by setting $n_0$ equal to $N(0)\Delta$,
the maximum quasiparticle density in the superconducting state. We
can evaluate $\lambda_{min}$ in the dirty s-wave case, as Kaplan
et al. \cite{kaplan} have calculated $\gamma_{pc}$ as well as
$\beta$ in this regime.  The characteristic phonon decay rate is
given by $\gamma_{pc}=4\pi^2N(0)\alpha^2\Delta/\hbar N$, where $N$
is the number density of ions.  Given this phonon decay rate, we
find that

\begin{equation} \lambda_{min}={2\pi\over\hbar}{N(0)\over NF(2\Delta)}. \label{eq:third} \end{equation}
The simplicity of this result is a consequence of detailed balance
in the dirty-limit.  Eq. 12 shows that the strength of the
bottleneck is determined by the ratio of the electron density of
states at the Fermi level to the phonon density of states at
2$\Delta$.  It is clear that the high-$T_c$ cuprates can be in the
weak bottleneck regime, where $\lambda$ is very small, in
constrast with conventional low-$T_c$ superconductors.  In the
high-$T_c$ materials the gap energy is large, and $2\Delta$
coincides with large phonon density of states peaks associated
with optic phonons.  Furthermore, the density of states at the
Fermi level is small compared with conventional metals.  Both
factors favor small $\lambda$. In contrast, in conventional
superconductors phonons with energy $2\Delta$ lie in the
low-energy tail of the acoustic spectrum.  Therefore the density
of resonant phonon states is small. This fact, together with the
large electronic density of states, place conventional
superconductors in the strong bottleneck regime.

\section{\label{sec:level1}Summary}

The most significant observation reported in this paper is the
strong dependence of the decay rate of the transient reflectivity
on both temperature and pump laser intensity.  At low $T$ the
decay rate is a linear function of $I$. At low $I$ the decay rate
decreases exponentially with decreasing $T$. At the lowest values
of $T$ and $I$ probed by our experiments, the lifetime of the
photoexited state becomes extremely long, $\approx 600$ ps. There
is no indication that the lifetime would not continue to increase
if the experiment could access lower values of $T$ and $I$.

In our analysis of the data, we first addressed the nature of the
excitations probed by measuring $\Delta R$.  We presented
arguments that $\Delta R$ is proportional to the density of
high-energy or antinodal quasiparticles created by the laser
excitation.  We explained theoretically why such particles,
although not the lowest energy excitations, could be highly
metastable at low temperature and density.  The metastability is a
consequence of the limited phase space for decay of high-energy
quasiparticles to the nodal regions of the Brillouin zone.  As the
phase space restrictions rely on momentum conservation, the degree
of metastability may be expected to be highly sensitive to
disorder.

Having understood the stability of an isolated quasiparticle, we
sought a quantitative understanding of the decay process that
occurs when a pair of quasiparticles interact. We estimated the
magnitude of the recombination coefficient, $\beta$, from both the
$T$ and $I$ dependence of the decay rate.  We emphasized that
$\beta$ is a direct measure of the strength of quasiparticle
interactions.  In addressing the theoretical approaches to the
recombination rate, we identified $\hbar/m^\ast$ as the natural
unit for $\beta$ in a two-dimensional superconductor.  We pointed
out that while recombination is assumed to be phonon-mediated in
conventional superconductors, electron-electron mediated processes
must be considered as well in the cuprates.  In the context of
phonon-mediated processes, we discussed the need to extend the
work of Kaplan et al. \cite{kaplan} to the 2D, clean limit that is
applicable to the cuprates.  In the electron-electron mediated
case, more work is needed to understand the underlying
recombination process. As a motivating factor, we pointed out the
remarkable coincidence between the normal state scattering rate
and the superconducting state recombination rate.  We showed that
both processes have the same dependence on the quasiparticle
density, which can be described by an interaction coefficient
equal to $\hbar/\pi m^\ast$.

This work was supported by NSF-9870258, DOE-DE-AC03-76SF00098,
CIAR and NSERC.

\end{document}